\documentclass[a4paper,fleqn]{cas-sc}

\usepackage[numbers]{natbib}
\usepackage{amsmath,scalerel}

\DeclareMathOperator*{\concat}{\scalerel*{\Vert}{\sum}}
\usepackage{float}

\def\tsc#1{\csdef{#1}{\textsc{\lowercase{#1}}\xspace}}
\tsc{WGM}
\tsc{QE}
\tsc{EP}
\tsc{PMS}
\tsc{BEC}
\tsc{DE}

\begin{document}
\let\WriteBookmarks\relax
\def\floatpagepagefraction{1}
\def\textpagefraction{.001}

\shorttitle{A Semi-Supervised Graph Autoencoder for Overlapping Community Detection}

\shortauthors{A Bekkair et~al.}

\title [mode = title]{A Noise-Resilient Semi-Supervised Graph Autoencoder for Overlapping Semantic Community Detection}


%


\author[]{Abdelfateh Bekkair}[orcid=0009-0003-8067-0580]

\cormark[2]


\ead{bekkair.abdelfateh@univ-ghardaia.edu.dz}

\affiliation[]{organization={Laboratoire des Mathématiques et Sciences Appliquées (LMSA), Université de Ghardaia},
    city={Ghardaia},
    country={Algeria}}

\author[]{Slimane Bellaouar} [orcid=0000-0001-8357-5501]
\cormark[1]

\ead{bellaouar.slimaneg@univ-ghardaia.edu.dz}
\author[]{Slimane Oulad-Naoui}[orcid=0000-0001-6311-5081]
\cormark[1]

\ead{s.ouladnaoui@univ-ghardaia.edu.dz}

\affiliation[]{organization={Faculty of Sciences and Technology, Université de Ghardaia},
    city={Ghardaia},
    country={Algeria}}

\cortext[cor1]{Corresponding author}
\cortext[cor2]{Principal corresponding author}

\begin{abstract}
Community detection in networks with overlapping structures remains a significant challenge, particularly in noisy real-world environments where integrating topology, node attributes, and prior information is critical. To address this, we propose a semi-supervised graph autoencoder that combines graph multi-head attention and modularity maximization to robustly detect overlapping communities. The model learns semantic representations by fusing structural, attribute, and prior knowledge while explicitly addressing noise in node features. Key innovations include a noise-resistant architecture and a semantic semi-supervised design optimized for community quality through modularity constraints.
Experiments demonstrate superior performance the model outperforms state-of-the-art methods in overlapping community detection (improvements in NMI and F1-score) and exhibits exceptional robustness to attribute noise, maintaining stable performance under 60\% feature corruption. These results highlight the importance of integrating attribute semantics and structural patterns for accurate community discovery in complex networks.
\end{abstract}








\maketitle

\section{Introduction}\label{intro}
Nodes in networks form densely connected groups, known as communities, where nodes within a community have more connections to each other than to nodes in other groups. Semantically, nodes in the same community share similar properties, roles, or functionalities~\citep{Fortunato2010}. Community detection in network analysis aims to uncover these communities and identify the functionalities of related nodes, providing insights into the underlying patterns and influences that shape real-world phenomena. 

The understanding and detection of communities have evolved over time, leading to the development of various techniques. These include hierarchical clustering~\citep{PhysRevE.69.066133,NewmanGirvan2004}, spectral clustering~\citep{de2014laplacian}, statistical inference~\citep{airoldi2008mixed}, label propagation~\citep{PhysRevE.76.036106}, graph representation through random walks\citep{grover2016node2vec}, and advanced approaches like graph autoencoders~\citep{10893950,SurveyAcomp}. Community detection in networks is a challenging endeavor, particularly with overlapping communities, as real-world networks often exhibit multiple memberships for nodes. This complexity not only expands the solution space but also reveals diverse associations among nodes, highlighting the intricate relationships within various network types.

Numerous algorithms for overlapping community detection have been proposed~\citep{agm,palla2005uncovering,yang2015defining,Dfuzzy,sci,WSCDSM,PSSNMTF,NMFjGO,SSGCAE}, with classical methods like Clique Percolation~\citep{palla2005uncovering} identifying communities through overlapping cliques. However, they suffer from high computational complexity and difficulty in detecting communities with sparse connectivity.
Graph Neural Networks (GNNs), such as the Neural Overlapping Community Detection (NOCD) model~\citep{nocd}, employ Graph Convolutional Networks and statistical inference but face limitations in detecting overlapping structures. The Self-Supervised Graph Convolutional Autoencoder (SSGCAE)~\citep{SSGCAE} combines graph autoencoder with modularity maximization, yet struggles with noisy attributes and insufficient semantic information, limiting its practical applicability.

In this paper, a semi-supervised model of graph autoencoder is presented for overlapping community detection based on modularity maximization through semantic information. This model combines topological information, attributed information, and prior information.

The performance of the proposed model is evaluated on three overlapping attributed networks: Facebook, Computer science, and Engineering against competitive models. Moreover, to assess our model's robustness to noise, we conduct a noise sensitivity analysis.

The paper is organized as follows: Section \ref{basic} provides a clear problem statement and outlines the employed evaluation metrics. Section \ref{relatedWork} delves into relevant research in the domain of non-overlapping and overlapping community detection. Section \ref{prop} presents our proposed methodology. Section \ref{exp} details the experimental setup, results, and discussion, including a parameter sensitivity analysis and noise robustness evaluation. Finally, Section \ref{conclusion} summarizes our findings and identifies potential directions for future research.

\section{Preliminaries}\label{basic}
This section introduces key concepts essential for analyzing complex networks: attributed networks, community detection, prior information integration, and evaluation metrics.

An attributed graph is defined by a set of nodes, a set of edges representing connections between these nodes, and attribute or feature information. This graph can be represented as a triplet \( G = \langle V, E, X \rangle \), where \( V = \{ v_1, v_2, \ldots, v_N \} \) denotes a set of \( N\) nodes, connected by a set of \( M \) edges \( E = \{ e_1, e_2, \ldots, e_M \} \). Each edge \(e_k\) is defined as a pair \(e_k = (v_i, v_j)\). The attribute matrix \( X = \{ x_1, x_2, \ldots, x_N \} \in \mathbb{R}^{N \times m} \) contains node feature information, with each vector \( x_i \) representing the attributes associated with node \( v_i \). The graph’s topological structure is captured by an adjacency matrix \( A \in \mathbb{R}^{N \times N} \), where each element is defined as follows:
\[
A_{ij} = \begin{cases}
1 & \text{if } (v_i, v_j) \in E, \\
0 & \text{otherwise.}
\end{cases}
\]

A community in a network is a group of nodes that are more densely connected with each other than with the rest of the network~\citep{PhysRevE.69.066133}. Community detection seeks to partition a graph \( G \) into \( k \) communities \( C = \{ C_1, C_2, \dots, C_k \} \). In the case of non-overlapping community detection, each community is a distinct subset of nodes, resulting in mutually exclusive groups where \( C_i \cap C_j = \emptyset \) for \( i \neq j \). However, many networks exhibit overlapping community structures, wherein nodes can simultaneously belong to multiple communities, making it necessary to allow intersections between communities such that \( C_i \cap C_j \neq \emptyset \) for some \( i \neq j \). Overlapping community detection thus aims to capture these shared memberships, providing a more nuanced representation of networks where nodes fulfill multiple roles or participate in multiple groups.

Semi-supervised methods leverage prior information, such as known node-community labels, to guide community detection. This approach combines labeled and unlabeled data to improve accuracy, often encoding prior knowledge in a matrix \( Y = [Y_{iq}] \), where:
\[
Y_{iq} = 
\begin{cases} 
1 & \text{if node } v_i \text{ belongs to community } C_q \\ 
0 & \text{otherwise} 
\end{cases}
\]
This prior matrix helps the model focus on both observed and inferred community structures.

To evaluate the detected communities, we use two commonly employed metrics for overlapping community structures: Overlapping Normalized Mutual Information (ONMI) and F1 score.

ONMI is the  extent of NMI for overlapping communities used to assess the similarity between detected and ground-truth community structures in overlapping networks~\citep{onmi}. It quantifies the amount of shared information between the two sets of communities, adjusted for overlap, with higher ONMI values indicating a closer match between predicted and true community structures.

The F1 score evaluates the accuracy of community detection by calculating the harmonic mean of precision and recall~\citep{f1}. This metric captures the balance between correctly identified community memberships and false positives or false negatives, providing a comprehensive measure of detection performance, particularly valuable in networks with overlapping structures.
\section{Related work}\label{relatedWork}

Numerous surveys and comparative studies have been conducted on community detection~\citep{jin2021survey,SurveyAcomp,Abekkair} and overlapping community detection~\citep{ding2016overlapping,vieira2020comparative,gupta2022review}. 

Classical approaches to soft community detection, such as Clique Percolation~\citep{palla2005uncovering}, Label Propagation~\citep{OLPA}, and Fuzzy Clustering~\citep{su2014quadratic} capture overlapping structures in networks by linking adjacent cliques, allowing multi-labeling of nodes and assigning membership degrees, respectively. Each method balancing between computational efficiency and interpretative complexity.
In statistical approaches, probabilistic models assign probabilistic affiliations to nodes to capture overlapping community memberships~\citep{airoldi2008mixed}. Bayesian methods update prior beliefs about community memberships based on observed data to handle uncertainty~\citep{CHEN2017790}. Whereas matrix factorization techniques decompose adjacency matrices into latent features to reveal shared memberships and uncover complex community structures~\citep{bigclam,sci,NMFjGO}.

Recent advances in deep learning have led to the development of several models for overlapping community detection. NOCD~\citep{nocd} integrates a Bernoulli–Poisson (BP) model with a two-layer Graph Convolutional Network (GCN) to derive community affiliation vectors, refining community structures by minimizing BP's negative log-likelihood and filtering weak affiliations. Alternatively, CommunityGAN~\citep{communityGAN} employs a generator-discriminator framework that uses nonnegative factors for node-community pairs and optimizes motif-level node preference distributions, enabling effective detection of overlapping communities. Additionally, the (Deep Learning-based Fuzzy Clustering Model) DFuzzy~\citep{Dfuzzy} adapts sparse networks through parallel processing to large-scale, using a stacked sparse autoencoder to focus on key nodes and evolve overlapping and disjoint communities, achieving significant improvements over traditional methods.

Several models incorporate prior information, such as WSCDSM~\citep{WSCDSM} recovers soft communities by learning a membership matrix within a unified non-negative matrix factorization (NMF) framework that integrates various inputs. In contrast, the semi-supervised models PSSNMTF~\citep{PSSNMTF} and SSGCAE~\citep{SSGCAE} adopt different strategies. PSSNMTF employs an NMF tri-factorization framework that incorporates node popularity, while SSGCAE leverages graph convolutional network (GCN) operations in conjunction with modularity maximization.

\section{Proposed model}\label{prop}
\begin{figure}[h!]
    \centering
    \includegraphics[width=0.7\linewidth]{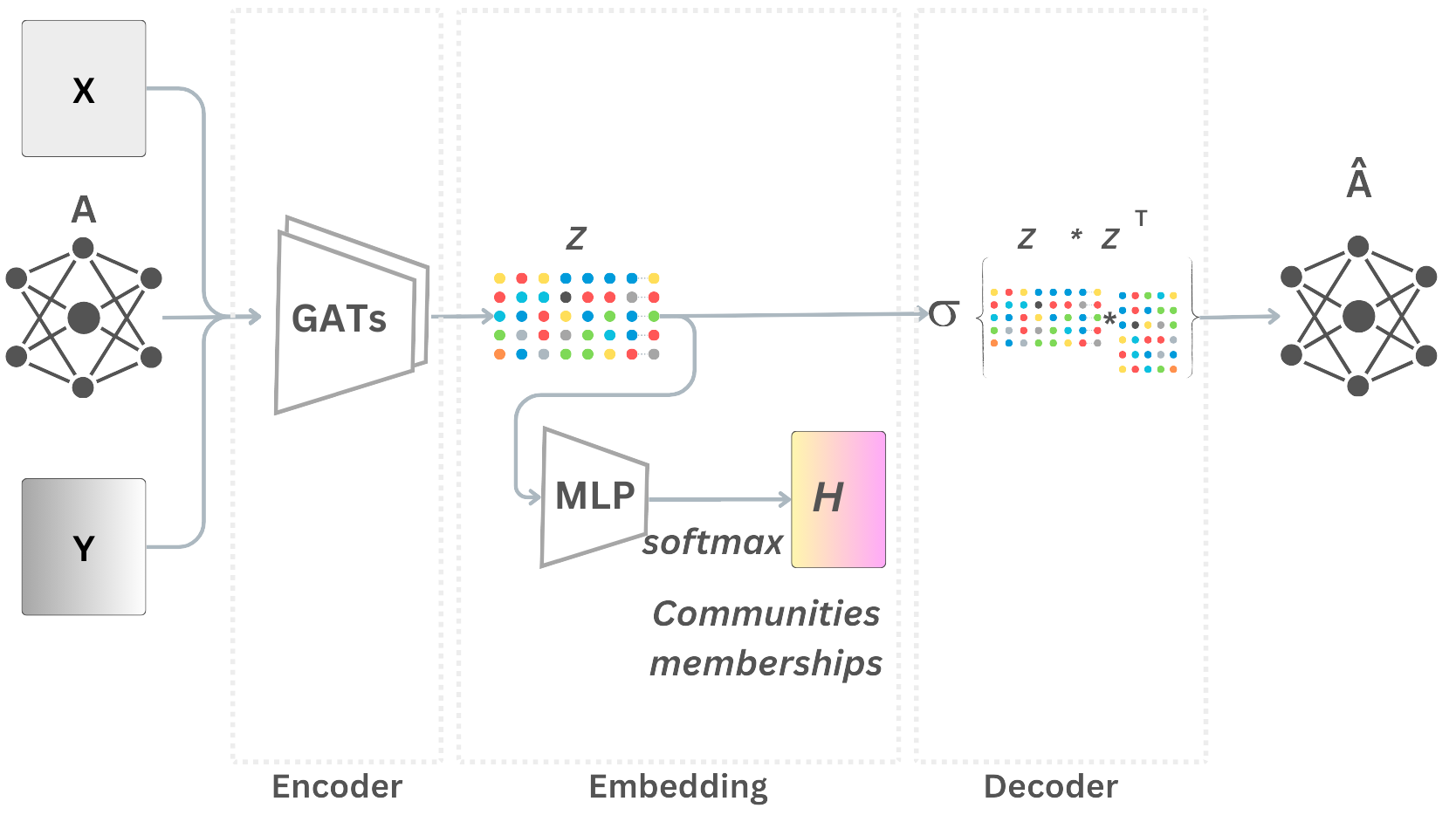}
    \caption{Architecture of the proposed model for overlapping community detection. Our model integrates attribute and prior information using a graph multi-attention autoencoder based on modularity maximization.}
    \label{fig:model}
\end{figure}

In this section, we introduce our proposed model, illustrated in Figure~\ref{fig:model}, and provide a detailed explanation of its key components: the encoder, the decoder, the employed objective functions, and the mechanism to assign community memberships.

Graph Attention Networks (GATs) we use to extract representations from topological information and attributed information~\citep{gat}. This choice is motivated by the ability of multi-head attention to capture richer semantic information compared to self attention techniques. To encode node representations, we utilize two layers of GAT, as shown in Equation~\ref{GAT}.
\begin{equation}
    z_{i} = \concat_{r=1}^r \sigma \left( \sum_{j \in \mathcal{N}_i} \alpha_{ij}^{r} W^r z_{j}^{(l-1)} \right)
    \label{GAT}
\end{equation}

In Equation~\ref{GAT}, \(\concat\) denotes concatenation across \( r \) attention heads, \( \sigma \) is the activation function, \(\mathcal{N}_i \) is the set of neighbors for node \( i \), and \( W^r \) is a learnable weight matrix for each attention head \( r \). The attention coefficient \( \alpha_{ij} \) which controls the influence of node \( j \) on node \( i \), is computed by Equation~\ref{att_coef}.

\begin{equation}
\alpha_{ij}^r = \frac{\exp\left(\text{LeakyReLU}\left(a^r(W^r h_i, W^r h_j)\right)\right)}{\sum_{m \in \mathcal{N}_i} \exp\left(\text{LeakyReLU}\left(a^r(W^r h_i, W^r h_m)\right)\right)}
 \label{att_coef}
\end{equation}
Here, \( h_i \) and \( h_j \) are the feature vectors of nodes \( i \) and \( j \), and \( a^r \) represent the attention mechanism that aggregates these features. This edge-specific compatibility score, weighted by the learnable matrix \( W^r \), enables the model to assign varying levels of influence to neighboring nodes.

The MLP acts as an additional layer applied to the embedding matrix \( \mathbf{Z} \in \mathbb{R}^{N \times d} \), where \( d \) is the embedding dimension. It performs a linear transformation parameterized by the weight matrix \( \mathbf{W} \) and the bias vector \( \mathbf{b} \). The output is then normalized using a softmax activation function to produce a probability distribution over \( k \) output communities. Formally, this process is given by \( \mathbf{H} = \text{Softmax}(\mathbf{WZ} + \mathbf{b}) \), where \( \mathbf{H} \in \mathbb{R}^{n \times k} \) contains the predicted community membership probabilities.

The encoded representations capture diverse compressed information about the network integrating both topological structure and attributed data. To decode and reconstruct the topological information, we apply an inner product to the embeddings \( Z \), enabling adjacency reconstruction, as shown in Equation~\ref{decoder}.
\begin{equation}
\hat{A} = \text{Sigmoid}(Z Z^T)
\label{decoder}
\end{equation}
This formulation uses the sigmoid function to generate a probability-based adjacency matrix, capturing the likelihood of connections between nodes in the reconstructed network.

The loss function in our model is designed with multiple objectives, specifically tailored to guide the model toward community detection and semi-supervised learning rather than focusing solely on reconstruction. As defined by Equation~\ref{loss}.
\begin{equation}
     \mathcal{L}= \mathcal{L}_{r}+ \alpha \mathcal{L}_{s} -\beta  \mathcal{L}_{c} 
\label{loss}
\end{equation}

Here, $\mathcal{L}_{r}$, $\mathcal{L}_{c}$, and $\mathcal{L}_{s}$ represent the reconstruction loss,  semi-supervised loss, and modularity maximization loss, respectively, with   $\alpha$ and $\beta$ as hyperparameters. Each function is detailed in the following parts.

As an autoencoder, our model aims to reconstruct the input from the encoded information. To quantify the reconstruction error, the reconstructed output is compared with the original input using a loss function. Specifically, we reconstruct the topological information by employing a binary cross-entropy loss function (Equation~\ref{binary_cross_loss}).

\begin{equation}
    \mathcal{L}_{r} = -\frac{1}{N} \sum_{i=1}^{N} \left[ A_i \cdot \log(\hat{A}_i) + (1 - A_i) \cdot \log(1 - \hat{A}_i) \right]
    \label{binary_cross_loss}
\end{equation}

To effectively guide the model in the graph clustering task, we utilize modularity maximization as the clustering loss function. This approach promotes the formation of densely connected intra-cluster nodes while minimizing connections between clusters. By maximizing modularity, as defined in Equation~\ref{modularityMAxLoss}, we achieve an optimized partitioning of the graph. This concept was first introduced by Newman and Girvan~\citep{NewmanGirvan2004}, it is based on the idea of comparing the actual edge density within a subgraph to the expected density from a null model. The latter is a random graph that preserves the original graph's degree distribution but lacks community structure, helping to identify statistically significant clusters.

\begin{equation}  \mathcal{L}_{c} =  \mathrm{Tr}(Z^\top \mathbf{B} Z)
    \label{modularityMAxLoss}
\end{equation}

Where, \( Z \) denotes the node embedding and \( \mathbf{B} \) represents the modularity matrix, as defined in Equation~\ref{modularity}. The trace operator \( (\text{Tr}) \) computes the sum of the diagonal elements, capturing the total contribution of assigned communities to the modularity score.
\begin{equation}
\mathbf{B} = \frac{1}{2m}\sum_{i,j}^n \left( A_{ij} - \frac{d_i d_j}{2m} \right) \delta(c_i, c_j)
    \label{modularity}
\end{equation}

In equation~\ref{modularity}, \( A_{ij} \) is the adjacency matrix, while \(d_i \) and \( d_j \) represent the degrees of nodes \( i \) and \( j \), respectively. The variable \( m \) denotes the total number of edges in the graph, and \( \delta(c_i, c_j) \) is an indicator function that equals 1 if nodes \( i \) and \( j \) are in the same cluster and 0 otherwise.

Semi-supervised learning in community detection leverages partial node-community labels \( Y \) to improve clustering quality and generalization, using cross-entropy as defined in the Equation~\ref{loss_semi}.

\begin{equation}
     \mathcal{L}_{S} = - \frac{1}{|\mathcal{T}|} \sum_{i \in \mathcal{T}} \sum_{c=1}^{C} y_{i,c} \log p_{i,c} 
     \label{loss_semi}
\end{equation}
\( \mathcal{L}_{S} \) calculates the average difference between the true labels \( y_{i,c} \) and predicted probabilities \( p_{i,c} \) for labeled nodes in the training set \( \mathcal{T} \). Here, \( y_{i,c} \) indicates the true class membership, while \( p_{i,c} \) represents the predicted likelihood for each class, aligning the model's predictions with known labels.

The community memberships of nodes are determined using a widely adopted strategy in soft clustering for community detection~\citep{startegySoftClustering}, based on a threshold \( \zeta \). If \( Z_{ip} \geq \zeta \), the node \( v_i \) is assigned to community \( c_p \). The threshold \( \zeta \) is calculated using Equation~\ref{zeta}~\citep{bigclam}.

\begin{equation}
\zeta=\sqrt{-\log\left(1- \frac{2m}{n(n-1)} \right)}
    \label{zeta}
\end{equation}
Where n and m are the number of nodes and edges, respectively.
\section{Experiments}\label{exp}
This section details our experimental setup. It include the used datasets, the noise robustness analysis, and a discussion of the results.

\subsection{Environment and tools}
The experiments were conducted on Kaggle\footnote{\url{www.kaggle.com}} platform utilizing 16 GB of RAM and a GPU featuring 15 GB of memory. We employed the \emph{PyTorch Geometric} library\footnote{\url{https://pytorch-geometric.readthedocs.io}}, a prominent tool in geometric deep learning, which provides essential features for managing graph-structured data and supports the development of GNN's efficiently. The source code is publicly available.\footnote{\url{https://github.com/abdelfateh10/SSGOCD}}
\subsection{Datasets}
We conduct experiments on three real-world network datasets, which include both social networks and co-authorship graphs. These datasets are briefly described in Table~\ref{tab:datasets}, highlighting key characteristics such as the number of nodes, edges, and community structure. Each dataset represents complex, real-world relationships where overlapping communities are present, providing a suitable environment for evaluating the performance of our model.

\begin{table}[h!]
\caption{Characteristics of the three attributed overlapping networks, k is the number of communities in each dataset.}
\label{tab:datasets}
\resizebox{0.6\columnwidth}{!}{%
\begin{tabular}{|l|l|l|l|l|l|}
\hline
\textbf{Dataset} & \textbf{Type of graph} & \textbf{\# Nodes} & \textbf{\# Edges} & \textbf{\# Features} & \textbf{k} \\ \hline
Facebook 1684    & Social        & 792    & 28,048  & 15    & 17 \\ \hline
Computer science & Co-authorship & 21,957 & 193,500 & 7,800 & 18 \\ \hline
Engineering      & Co-authorship & 14,927 & 98,610  & 4,800 & 16 \\ \hline
\end{tabular}%
}
\end{table}
\subsection{Settings}
The model employs eghit attention heads at the two layers of the encoder, with a hidden layer  of \( 64 \times 8 \) units and an embedding size of \( 16 \times 8 \). The size of the membership matrix is determined by the number of communities specific to each dataset. The learning rate is set to \( 0.006 \), while the coefficients $\alpha$ and $\beta$ are fixed at \(  0.5 \) and \( 1 \times 10^{-6} \), respectively. All methods were run ten times, and the average results are reported. 
\subsection{Experimental Analysis and Discussion}
In our research, we evaluate the efficiency and robustness of the proposed approach through experiments on overlapping community detection, both in the absence of noise and under noisy condition. In the sequel we detail of these experiments are provided.

\subsubsection{Experiment on Overlapping Community Detection}
The experimental results on overlapping community detection without noisy for Facebook1682, Engineering, and Computer Science datasets are shown in Tables~\ref{tab:onmi}, and~\ref{tab:f1}, respectively. As can observe that the three models SCI, NMFJGO, and NOCD demonstrate consistent performance, unaffected by variations in the percentage of labeled input. This is expected a behavior since these models are not designed to leverage unlabeled data, unlike the other semi-supervised model. 
WSCDSM and PSSNMTF show minor improvements but remain less competitive. In contrast, both SSGCAE and our model excel in ONMI and F1 metrics. For ONMI, Our model consistently achieves the highest scores across all datasets, particularly at 10\% highlighting its adaptability and effectiveness in leveraging additional labeled data. Similarly, in terms of F1 scores, our model significantly outperforms all others, achieving the best scores at 10\%: 0.8 for FB1684, 0.86 for Eng, and 0.82 for CS. These results demonstrate the model's robust performance and superior ability to utilize additional labeled data, showcasing dramatic improvements in both ONMI and F1 scores compared to the unlabeled scenario. This underscores the model's effectiveness in community detection task across diverse datasets. High ONMI and F1 scores indicate that the model effectively detects community structures and classifies nodes accurately. As more labeled data becomes available, our model consistently achieves the best performance, demonstrating its strength in community detection.
\begin{table}[!ht]
\caption{Performance on ONMI across the three datasets.}
\label{tab:onmi}
\resizebox{0.5\linewidth}{!}{%
\begin{tabular}{|l|l|l|l|l|l|l|}
\hline
\textbf{Dataset} & \textbf{Model}    & \textbf{02\%} & \textbf{04\%} & \textbf{06\%} & \textbf{08\%} & \textbf{10\%} \\ \hline
\multirow{7}{*}{FB1684} 
                 & \textbf{SCI}       & 0.23 & 0.23 & 0.23 & 0.23 & 0.23 \\ \cline{2-7} 
                 & \textbf{NMFjGO}    & 0.10 & 0.10 & 0.10 & 0.10 & 0.10 \\ \cline{2-7} 
                 & \textbf{NOCD}    & 0.27 & 0.27 & 0.27 & 0.27 & 0.27 \\ \cline{2-7} 
                 & \textbf{WSCDSM}    & 0.24 & 0.23 & 0.26 & 0.25 & 0.30 \\ \cline{2-7} 
                 & \textbf{PSSNMTF}   & 0.15 & 0.17 & 0.26 & 0.22 & 0.24 \\ \cline{2-7} 

                 & \textbf{SSGCAE}   & \textbf{0.42}& 0.43 & 0.47 & 0.47 & 0.48 \\ \cline{2-7} 
                 & \textbf{Our}   &   0.40  & \textbf{0.46}  & \textbf{0.50}  & \textbf{0.52} & \textbf{0.54}\\ \hline

\multirow{7}{*}{Eng} 
                 & \textbf{SCI}       & 0.10 & 0.10 & 0.10 & 0.10 & 0.10 \\ \cline{2-7} 
                 & \textbf{NMFjGO}    & 0.20 & 0.20 & 0.20 & 0.20 & 0.20 \\ \cline{2-7} 
                 & \textbf{NOCD}      & 0.25 & 0.25 & 0.25 & 0.25 & 0.25 \\ \cline{2-7} 
                 & \textbf{WSCDSM}     & 0.11 & 0.12 & 0.07 & 0.12 & 0.11 \\ \cline{2-7} 
                 & \textbf{PSSNMTF}    & 0.16 & 0.16 & 0.20 & 0.18 & 0.20 \\ \cline{2-7} 
                 & \textbf{SSGCAE}   &   0.57 & 0.61 & 0.61 & 0.62 & 0.63 \\ \cline{2-7} 
                 & \textbf{Our}  &\textbf{0.60} & \textbf{ 0.62} &\textbf{0.64} &\textbf{0.64} & \textbf{0.65}\\ \hline

\multirow{7}{*}{CS} 
                 & \textbf{SCI}       & 0.09 & 0.09 & 0.09 & 0.09 & 0.09 \\ \cline{2-7} 
                 & \textbf{NMFjGO}   & 0.25 & 0.25 & 0.25 & 0.25 & 0.25 \\ \cline{2-7} 
                 & \textbf{NOCD}      & 0.35 & 0.35 & 0.35 & 0.35 & 0.35 \\ \cline{2-7} 
                 & \textbf{WSCDSM}   & 0.10 & 0.10 & 0.12 & 0.12 & 0.09 \\ \cline{2-7} 
                 & \textbf{PSSNMTF}    & 0.16 & 0.17 & 0.18 & 0.20 & 0.20 \\ \cline{2-7} 
                 & \textbf{SSGCAE}   &   0.42 & 0.52 & 0.53 & 0.53 & 0.54 \\ \cline{2-7} 
                 & \textbf{Our} &   \textbf{0.54} &\textbf{0.55}& \textbf{0.55} & \textbf{0.56} & \textbf{0.57} \\ \hline
\end{tabular}%
}
\end{table}
\begin{table}[!ht]
\caption{Performance on F1-measure across the three datasets.}
\label{tab:f1}
\resizebox{0.5\linewidth}{!}{%
\begin{tabular}{|l|l|l|l|l|l|l|}
\hline
\textbf{Dataset} & \textbf{Model}    & \textbf{02\%} & \textbf{04\%} & \textbf{06\%} & \textbf{08\%} & \textbf{10\%} \\ \hline
\multirow{7}{*}{FB1684} 
                 & \textbf{SCI}        & 0.14 & 0.14 & 0.14 & 0.14 & 0.14 \\ \cline{2-7} 
                 & \textbf{NMFjGO}    & 0.11 & 0.11 & 0.11 & 0.11 & 0.11 \\ \cline{2-7} 
                 & \textbf{NOCD}      & 0.14 & 0.14 & 0.14 & 0.14 & 0.14 \\ \cline{2-7} 
                 & \textbf{WSCDSM}     & 0.15 & 0.18 & 0.18 & 0.20 & 0.15 \\ \cline{2-7} 
                 & \textbf{PSSNMTF}   & 0.13 & 0.11 & 0.14 & 0.14 & 0.18 \\ \cline{2-7} 
                 & \textbf{SSGCAE}   & 0.38 & 0.40 & 0.51 & 0.54 & 0.63 \\ \cline{2-7} 
                 & \textbf{Our} & \textbf{0.62} & \textbf{0.71} & \textbf{0.75} & \textbf{0.76} & \textbf{0.80}\\ \hline

\multirow{7}{*}{Eng} 
                 & \textbf{SCI}       & 0.12 & 0.12 & 0.12 & 0.12 & 0.12 \\ \cline{2-7} 
                 & \textbf{NMFjGO}    & 0.15 & 0.15 & 0.15 & 0.15 & 0.15 \\ \cline{2-7} 
                 & \textbf{NOCD}      & 0.09 & 0.09 & 0.09 & 0.09 & 0.09 \\ \cline{2-7} 
                 & \textbf{WSCDSM}    & 0.12 & 0.14 & 0.11 & 0.12 & 0.13 \\ \cline{2-7} 
                 & \textbf{PSSNMTF}    & 0.15 & 0.15 & 0.15 & 0.14 & 0.14 \\ \cline{2-7} 
                 & \textbf{SSGCAE}     & 0.45 & 0.50 & 0.50 & 0.54 & 0.60 \\ \cline{2-7} 
                 & \textbf{Our} &\textbf{0.84}&\textbf{0.85} &\textbf{0.86} & \textbf{0.86} & \textbf{0.86} \\ \hline

\multirow{7}{*}{CS} 
                 & \textbf{SCI}       & 0.07 & 0.07 & 0.07 & 0.07 & 0.07 \\ \cline{2-7} 
                 & \textbf{NMFjGO}    & 0.13 & 0.13 & 0.13 & 0.13 & 0.13 \\ \cline{2-7} 
                 & \textbf{NOCD}       & 0.22 & 0.22 & 0.22 & 0.22 & 0.22 \\ \cline{2-7} 
                 & \textbf{WSCDSM}     & 0.14 & 0.15 & 0.15 & 0.15 & 0.16 \\ \cline{2-7} 
                 & \textbf{PSSNMTF}    & 0.11 & 0.11 & 0.12 & 0.17 & 0.19 \\ \cline{2-7} 
                 & \textbf{SSGCAE}   & 0.44 & 0.50 & 0.55 & 0.65 & 0.68 \\ \cline{2-7} 
                 & \textbf{Our} &  \textbf{0.80} & \textbf{0.81} &\textbf{ 0.82} & \textbf{0.82} & \textbf{0.82}\\ \hline
\end{tabular}%
}
\end{table}

\subsubsection{Noise Robustness Analysis}
To evaluate our model's ability to maintain performance and stability under varying levels of data noise, we simulate noisy attribute information in this section by randomly exchanging attribute vectors of with a rate \( P_{\text{mis}} \) varying from 20\% to 60\%.

\begin{figure}[!ht]
    \centering
    \includegraphics[width=0.45\linewidth]{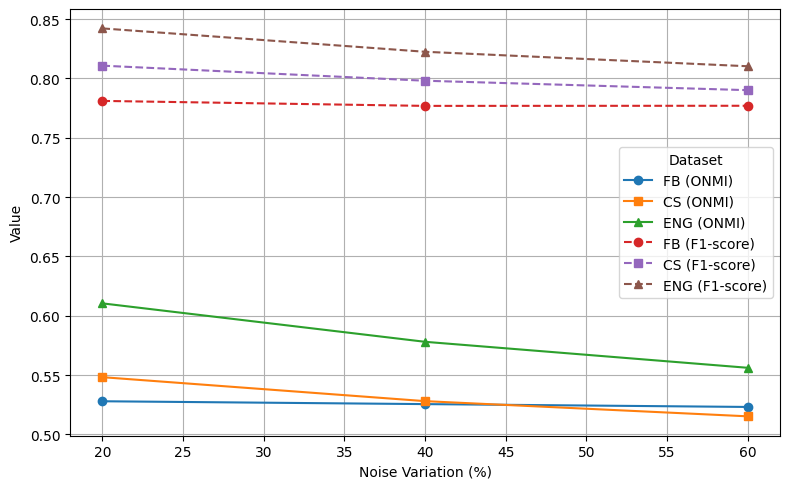}
    \caption{Sensitivity analysis of model performance on the Facebook, Engineering, and Computer Science datasets under varying noise levels ($P_{\mathrm{mis}}$), with prior information fixed at 10\%. Robustness was evaluated using ONMI and F1‑score to assess stability as $P_{\mathrm{mis}}$ increases.}

    \label{fig:NoiseVariation}
\end{figure}

Figure~\ref{fig:NoiseVariation} shows our model’s performance across varying levels of $P_{\mathrm{mis}}$, with the prior information rate fixed at 10\%, underscoring its robustness: despite increasing noise, the model remains stable. In the FB dataset, results drop by only 1\% for both ONMI and F1 at $P_{\mathrm{mis}}=60\%$ compared to $20\%$. Similarly, in the ENG dataset ONMI declines by 3\% and F1 by 2\% at $P_{\mathrm{mis}}=60\%$. In the CS dataset ONMI decreases by 6\% and F1 by 3\% under the same conditions. These findings confirm that our model maintains competitive performance even under high levels of noisy attribute information.

\begin{figure}[!ht]
    \centering
    \includegraphics[width=0.9\linewidth]{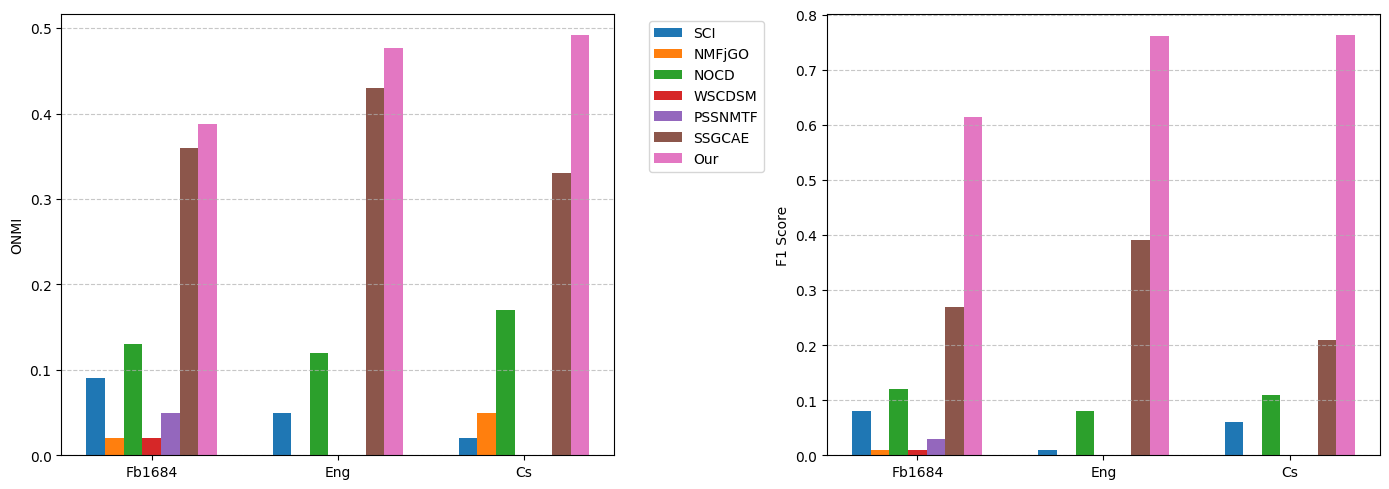}
\caption{Performance comparison of all studied models in terms of ONMI and F1-measure on the FB dataset, Engineering dataset, and Computer Science dataset, with \( P_{\text{mis}} \) fixed at 60\% noise attribution and 02\% of prior information.}
    \label{fig:noiseAnalysis}
\end{figure}

To compare our model with other studied models, Figure~\ref{fig:noiseAnalysis} presents the ONMI and F1 results on the FB1684, ENG, and CS datasets across all models considered in the experiment. The results indicate that most models do not achieve acceptable performance. However, both SSGCAE and our proposed model demonstrate significantly better results, standing out as the only models capable of achieving competitive performance.


\section{Conclusion}\label{conclusion}
To enhance the discovery of overlapping communities within networks, future efforts should focus on integrating semantic information by combining link and attribute data. Additionally, leveraging prior knowledge can improve model robustness, addressing challenges posed by sensitivity to noisy attributes. In this work, 
We have proposed a model for detecting overlapping communities that is robust to noisy attributes. The model leverages a semantic graph autoencoder, that integrate a graph attention encoder, a modularity maximization objective, and prior information. Experiments conducted on three overlapping networks, both with and without noisy attributes, demonstrate the efficiency and robustness of the proposed approach.

Future work will focus on extending the model to handle large-scale networks efficiently, enhancing its interpretability to provide clearer insights into overlapping community structures, and adapting it to dynamic networks where community structures evolve over time.

\section*{Acknowledgments}
\begin{sloppypar}This work is supported by the DGRSDT-Algeria, under the PRFU Project: C00L07UN470120230001. \end{sloppypar}
\printcredits
\bibliographystyle{splncs04}

\bibliography{cas-refs}
\end{document}